# Recurrence of galactic cosmic rays intensity and anisotropy in solar minima 23/24 and 24/25 by ACE/CRIS, STEREO, SOHO/EPHIN and neutron monitors – Fourier and wavelet analysis (Dedicated to the Memory of Michael Alania)

R. Modzelewska[1] and A. Gil[1,2]

[1] Faculty of Exact and Natural Sciences, Institute of Mathematics, Siedlce University, Konarski Str. 2 08-110 Siedlce, Poland,

[2] Space Research Centre, Polish Academy of Sciences, Bartycka Str.18A, 00-716Warsaw, Poland

**ABSTRACT**

Aims. We study the 27-day variations of galactic cosmic rays (GCRs) based on neutron monitor (NM), ACE/CRIS, STEREO and SOHO/EPHIN measurements, in solar minima 23/24 and 24/25 characterized by the opposite polarities of solar magnetic cycle. Now there is an opportunity to reanalyze the polarity dependence of the amplitudes of the recurrent GCR variations in 2007-2009 for negative $A < 0$ solar magnetic polarity and to compare it with the clear periodic variations related to solar rotation in 2017-2019 for positive $A > 0$.

Methods. We use the Fourier analysis method to study the periodicity in the GCR fluxes. Since the GCR recurrence is a consequence of solar rotation, we analyze not only GCR fluxes, but also solar and heliospheric parameters examining the relationships between the 27-day GCR variations and heliospheric, as well as, solar wind parameters.

Results. We find that the polarity dependence of the amplitudes of the 27-day variations of the GCR intensity and anisotropy for NMs data is kept for the last two solar minima: 23/24 (2007-2009) and 24/25 (2017-2019) with greater amplitudes in positive $A > 0$ solar magnetic polarity. ACE/CRIS, SOHO/EPHIN and STEREO measurements are not governed by this principle of greater amplitudes in positive $A > 0$ polarity. GCR recurrence caused by the solar rotation for low energy (< 1GeV) cosmic rays is more sensitive to the enhanced diffusion effects, resulting in the same level of the 27-day amplitudes for positive and negative polarities. While high energy (> 1GeV) cosmic rays registered by NMs, are more sensitive to the large-scale drift effect leading to the 22-year Hale cycle in the 27-day GCR variation, with the larger amplitudes in the $A > 0$ polarity than in the $A < 0$.

## 1. Introduction

The galactic cosmic ray (GCR) spectrum observed near the Earth is considerably controlled by the solar activity. Transport of GCRs in the heliosphere is governed by the solar wind and the combination of the regular and turbulent heliospheric magnetic field (HMF). Due to the interaction of GCRs with solar wind and HMF, GCR spectrum below a few tens of GV, is modulated with respect to the local interstellar spectrum (LIS) (Potgieter 2013). GCR flux is a subject of solar modulation producing various quasi-periodic changes in different time scales (from hours to several years), see e.g. (Usoskin 2017; Kudela & Sabbah 2016; Chowdhury et al. 2016; Bazilevskaya et al. 2014) and references therein. The most prominent periodic variations are the 22 years, 11 years, 27 days and solar diurnal variation (24 hours). On the background of the long term solar modulation: 22 years and 11 years, connected with the global HMF and the solar activity cycle, respectively, the short term modulation effects: sporadic (Forbush decreases) and recurrent (periodic) also take place.

The 27-day variation of the GCR intensity and anisotropy, related to the solar rotation, is the main topic of this paper. This phenomenon is connected with the heliolongitudinal asymmetry of the heliospheric parameters being under the influence of corotating interaction regions (CIRs), for review see (Richardson 2004, 2018). The Sun rotates at different rates depending on the heliolatitudes, which is called differential rotation. The equator and near equatorial regions of the Sun rotate on its axis with the period of about 25–26 days, called the Sun's sidereal period of rotation. For the observer on the Earth this periodicity equals 27–28 days due to the orbital motion of the Earth, called the Sun's sinodial period of rotation. At its polar regions the period value of the Sun rotation is about 36 days. The recurrent variations of GCR with the period of ∼ 27 days are generally more typical and longer in duration during the minimum and near minimum epochs of solar activity. The amplitudes of the 27-day variations, and other GCR variations (e.g. Forbush decreases and solar diurnal anisotropies) in the range between few hours and few months vary, in general, in phase with the solar activity cycle, i.e. they reach the maximum values during solar activity maximum (e.g., Bazilevskaya 2000). Recurrent structures in GCR fluxes with the period of ∼ 27 days are observed at the Earth by neutron monitors (NMs) (Modzelewska & Alania 2013), in space by e.g.

ACE (Leske et al. 2011), out of the ecliptic plane by Ulysses (Mckibben et al. 1995) and even in the distant heliosphere by the Voyager spacecraft (Decker et al. 1999).

Richardson et al. (1999) showed for the first time that the amplitude of the recurrent GCR variation is larger when $A > 0$ (A is the global direction of the HMF) than during adjacent minima when $A < 0$. Until now the polarity dependence of the amplitude of the 27-day variation of the GCR intensity was mainly considered quantitatively by NMs observations (Alania et al. 2001; Gil & Alania 2001; Vernova et al. 2003). It was also confirmed using mathematical modeling of GCR heliospheric transport (e.g., Iskra et al. 2004). Later, (Alania et al. 2005, 2008; Gil et al. 2005) showed that the amplitudes of the 27-day variation of the GCR anisotropy are also polarity dependent at solar minimum, with greater values when $A>0$ than in $A<0$ polarity period. The polarity dependence of the 27-day amplitude of GCR registered by NMs was confirmed experimentally lately by Gil & Mursula (2017), explaining it by combination of drift effect and solar wind convection. In this paper we extend the quantitative study of the 27-day variation of the GCR intensity for lower part of the energy spectrum using spacecraft data: SOHO/EPHIN, STEREO and ACE/CRIS. It is now an opportunity to re-analyze the polarity dependence of the amplitudes of the recurrent GCR variations in solar minimum 23/24 in 2007-2009 when $A < 0$ and to compare it with the clear periodic variations related to solar rotation in solar minimum 24/25 in 2017-2019 when $A > 0$. Since the GCR recurrence is a consequence of solar rotation, we analyze not only GCR fluxes, but also solar and heliospheric parameters examining the relationship between the 27-day GCR variations and heliospheric, as well as, solar wind parameters.

The paper is organized as follows: in Section 2 we give the description of the analysed data and methods used in this paper. In Section 3 we study the polarity dependence of the amplitudes of the 27-day GCR variations. In Section 4 we analyze the relationship of the periodic GCR variation with heliospheric parameters. In Section 5 we discuss the obtained results. In Section 6 we give our conclusions.

## 2. Data and methods

We analyse the daily data of GCR proton flux for SOHO/EPHIN, STEREO A and B for solar minimum 23/24 when $A < 0$ in 2007-2009 and solar minimum 24/25 when $A > 0$ in 2017-2019. STEREO B data are not available in 2017-2019. The same we perform for ACE/CRIS GCR fluxes of Carbon, Nitrogen, Oxygen, Neon, Silicon and Iron. Additionally we update the calculations for the daily GCR intensity and anisotropy for NMs with different cut off rigidities, from 0.67 GV to 4.00 GV: Hermanus, Newark, Kerguelen, Oulu and Apatity.

We study the dynamics of the recognized periodicity connected with the solar rotation using Lomb periodogram (Lomb 1976) adapting the procedure from (Modzelewska & Alania 2013). We calculate the power of the periodic variation connected with the Sun's rotation for period of $27_2$ days with 95% confidence level. Maximal power from Lomb periodogram indicates the main period of the recurrent GCR variations in the analysed time series. As an example Figure 1 (1st panel) presents STEREO A daily proton fluxes for kinetic energy 40-60 MeV and 60-100 MeV for solar minimum 23/24 when $A < 0$ in 2007-2009. The second panel of Figure 1 presents normalized and detrended (by excluding 29 days running average) GCR intensity. The timeline of the power of recurrent variations lasting 25-29 days and the recognized main period for each Bartel rotation (BR) are displayed on the third and fourth panels of Figure 1, respectively. Figure 2 presents data and Lomb analysis results of STEREO A for solar minimum 24/25 when $A > 0$ in 2017-2019. Figures 1 and 2 demonstrate the existence of the periodic variations of the GCR intensity related to solar rotation for the last two solar minima with opposite solar magnetic polarity: 23/24 in 2007-2009 and 24/25 in 2017-2019. One can observe the high power of the very famous episode of the 27-day variation when $A < 0$ in 2007-2008 (Fig. 1) and clear periodic variations when $A > 0$ in 2017-2018 (Fig. 2) for both energy bins. The same data processing was done for STEREO B, SOHO/EPHIN, ACE/CRIS and NMs data revealing similar results.

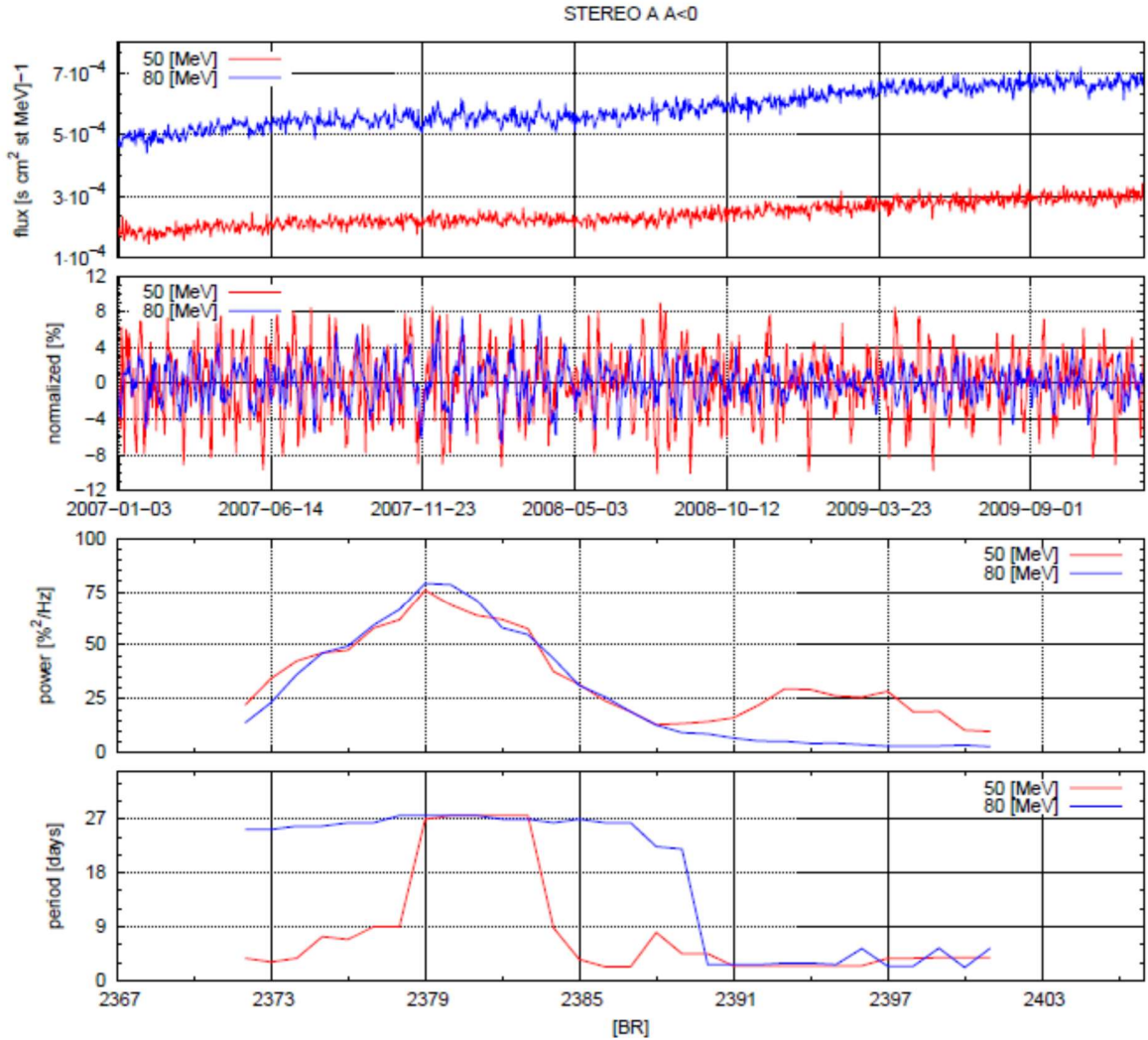

**Fig. 1.** (1st panel) *STEREO* A daily proton fluxes for kinetic energy 40-60 MeV and 60-100 MeV for solar minimum 23/24 when $A < 0$ in 2007-2009; (2nd panel) GCR intensity normalized and detrended by excluding 29 days running average; the timeline of the power of recurrent variations lasting 25-29 days (3rd panel) and the recognized main period (4th panel).

### 3. Polarity dependence of the amplitudes of the 27-day variations

First we study the polarity dependence of the amplitude of the 27-day variation using NMs data: Hermanus, Newark, Kerguelen, Oulu and Apatity. We update the calculations for the 27-day vari-

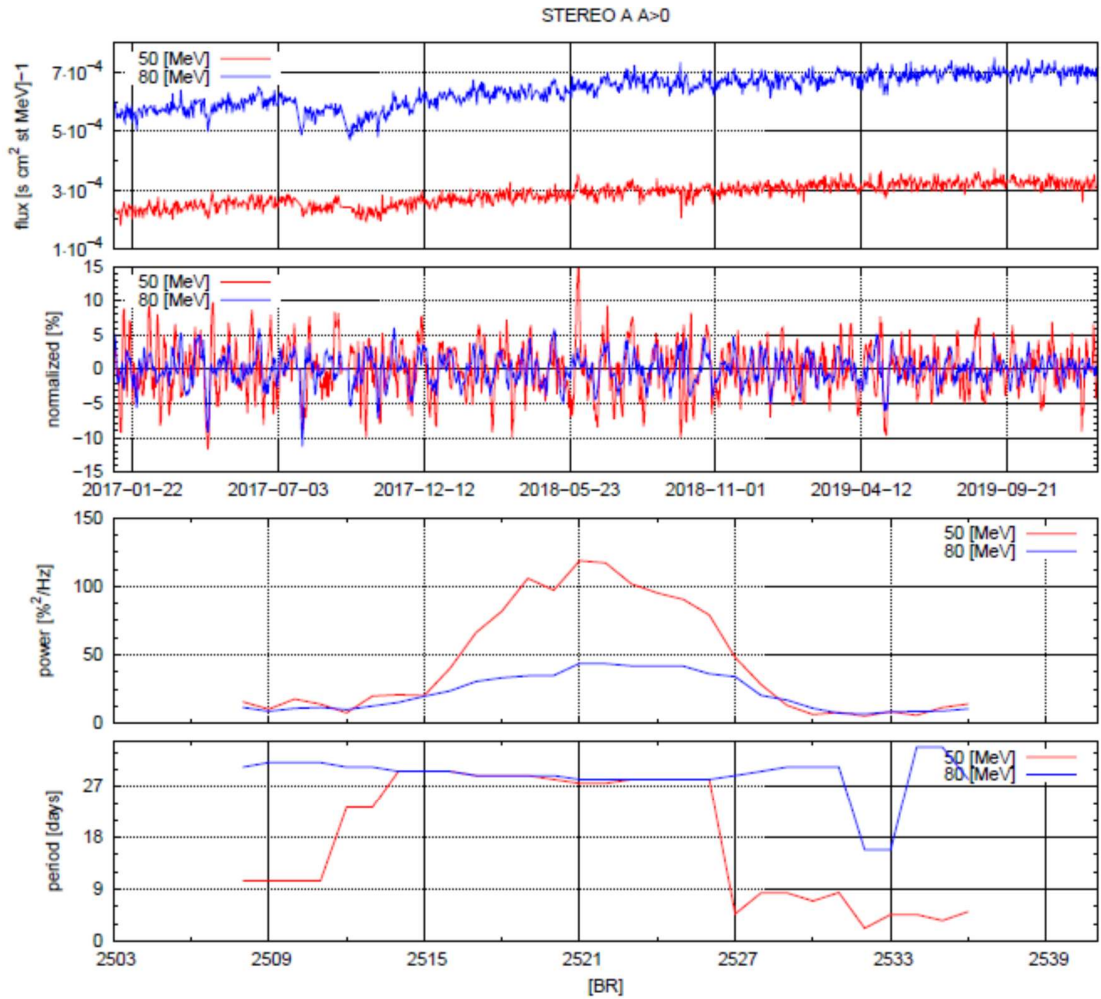

**Fig. 2.** The same as in Fig. 1 for solar minimum 24/25 when $A > 0$ in 2017-2019

ation of the GCR anisotropy A27A (Modzelewska & Alania 2018) and intensity A27I (Gil & Mursula 2017) for the solar minima 23/24 and 24/25. The average amplitudes of the A27A and A27I for the last two solar minima are presented in Figure 3 as well in Table 1. We find that the average A27A and A27I are polarity dependent, with greater amplitudes when $A > 0$ in 2017-2019 for solar minimum 24/25 than for $A < 0$ in 2007-2009 for solar minimum 23/24:

$A27A_{A>0} = 0:155 \_ 0:007$ [%]
$A27A_{A<0} = 0:125 \_ 0:004$ [%]
$A27I_{A>0} = 0:520 \_ 0:030$ [%]
$A27I_{A<0} = 0:350 \_ 0:030$ [%].

Figure 4 and Table 2 present the average A27 of the GCR proton intensity for SOHO/EPHIN (energy bins: 25-40.9 MeV/n and 40.9-53 MeV/n) and STEREO A, B (energy bins: 40-60 MeV/n and 60-100 MeV/n) for $A > 0$ and $A < 0$ polarity. Although, the picture in Figure 4 is quite complicated, the weak tendency of greater amplitudes for $A > 0$ than for $A < 0$ for higher energies 40-100 MeV/n in STEREO data can be seen, but opposite scenario is noticeable for lower energies 25-53 MeV/n in SOPHO/EPHIN data.

Figures 5 - 6 and Tables 3 - 4 present the A27 of GCR fluxes for Carbon, Nitrogen, Oxygen, Neon, Silicon and Iron by ACE/CRIS, respectively. For each species we analyse 7 energy bins

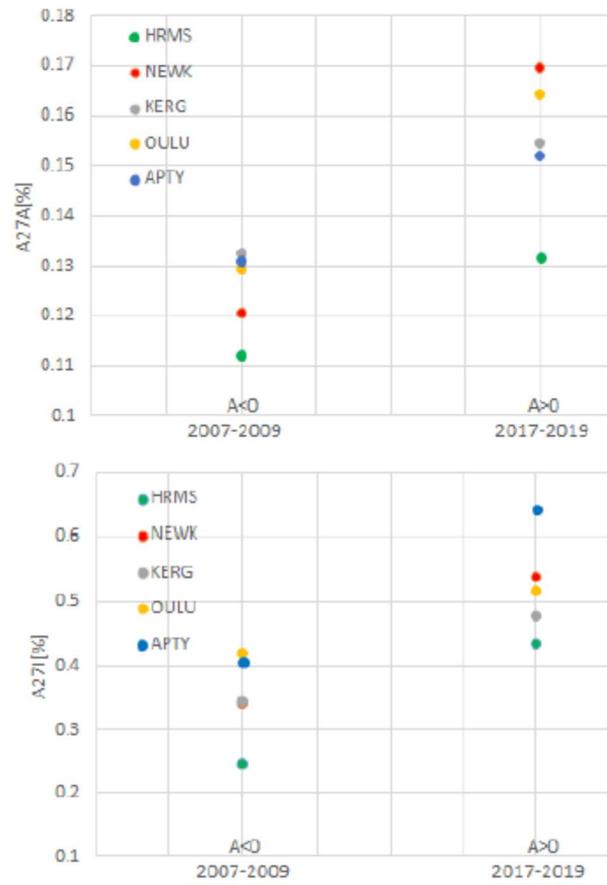

**Fig. 3.** Amplitude of the 27-day variation of the GCR anisotropy (top) and intensity (bottom) observed by NMs (Hermanus, Rc=4.58GV; Newark, Rc=2.40GV, Kerguelen, Rc=1.14GV; Oulu, Rc=0.81GV; Apatity, Rc=0.65GV) for solar minimum 23/24 when $A < 0$ in 2007-2009 and solar minimum 24/25 when $A > 0$ in 2017-2019.

**Table 1.** Amplitude of the 27-day variation of the GCR anisotropy (A27A) and intensity (A27I) observed by NMs for $A > 0$ and $A < 0$ polarities.

| A27A[%] NM station | $A < 0$ 2007-2009 | $A > 0$ 2017-2019 |
|---|---|---|
| Apatity | 0.13±0.01 | 0.15±0.01 |
| Kerguelen | 0.13±0.01 | 0.15±0.01 |
| Newark | 0.12±0.01 | 0.17±0.02 |
| Oulu | 0.13±0.01 | 0.16±0.02 |
| Hermanus | 0.11±0.01 | 0.13±0.01 |

| A27I[%] NM station | $A < 0$ 2007-2009 | $A > 0$ 2017-2019 |
|---|---|---|
| Apatity | 0.41±0.04 | 0.64±0.05 |
| Kerguelen | 0.34±0.04 | 0.48±0.05 |
| Newark | 0.34±0.04 | 0.54±0.07 |
| Oulu | 0.42±0.05 | 0.52±0.05 |
| Hermanus | 0.25±0.02 | 0.43±0.05 |

available at ACE/CRIS measurements. Values of the A27 for GCR flux of heavier nuclei registered by ACE/CRIS remain on the same level for both polarities for all considered species.

**4. Periodic GCR variations and heliospheric parameters**

We consider here daily data of solar and heliospheric parameters, such as sunspot number (SSN), solar radio flux (SRF), solar wind speed (SWs), temperature (SWT) and density (SWd), HMF strength (B) and components (Bx; By; Bz). First, we use cross-wavelet transform to trace an even-tual coherence between cosmic rays and the above mentioned time series around time-period of solar rotation. We perform the cross-wavelet spectrum and the wavelet coherence analysis for each of the above data in the pair with the GCR flux from Oulu NM, for both studied here solar minima. Crosswavelet analysis is built on the basis of continuous wavelet transform. It simply consists of two wavelet transforms. Signal is decomposed into scaled and shifted version of a wavelet being a function of two variables: translation and scale. Wavelet transform is localized in both, time and frequency domain. In our calculations we follow Torrence & Webster (1999) using tools given by Jevrejeva et al. (2003) and Grinsted et al. (2004). Comparing solar minimum 23/24 and 24/25 it is clear that in all of the studied parameters the 27-days variation was prevalent in both time intervals, although was more incessantly visible during the former one (compare, e.g., Modzelewska & Alania 2013; Gil & Mursula 2018; Jian et al. 2019).

**Table 2.** Amplitude of the 27-day variation of the GCR intensity for protons observed by *SOHO/EPHIN*, *STEREO* A and B for $A > 0$ and $A < 0$ polarities.

| A27I[%] E[MeV/n] | $A < 0$ 2007-2009 | $A > 0$ 2017-2019 |
|---|---|---|
| *SOHO/EPHIN* | | |
| 25-40.9 | 3.94±0.00 | 3.04±0.36 |
| 40.9-53 | 4.25±0.00 | 3.44±0.37 |
| *STEREO A* | | |
| 40-60 | 2.17±0.20 | 2.65±0.20 |
| 60-100 | 1.61±0.20 | 1.83±0.10 |
| *STEREO B* | | |
| 40-60 | 2.80±0.30 | —— |
| 60-100 | 1.40±0.10 | —— |

**Table 3.** Amplitude of the 27-day variation of the GCR intensity for Carbon, Nitrogen and Oxygen by *ACE/CRIS* for $A > 0$ and $A < 0$ polarities.

| A27I[%] ACE C E[MeV/n] | $A < 0$ 2007-2009 | $A > 0$ 2017-2019 | A27I[%] ACE N E[MeV/n] | $A < 0$ 2007-2009 | $A > 0$ 2017-2019 |
|---|---|---|---|---|---|
| 47.8-61.2 | 4.03±0.32 | 3.74±0.32 | 51.4-65.8 | 6.47±0.53 | 6.93±0.54 |
| 62.6-84.0 | 3.06±0.26 | 3.00±0.24 | 67.3-90.4 | 6.26±0.38 | 5.55±0.46 |
| 85.0-102.7 | 3.33±0.30 | 3.30±0.27 | 91.5-110.6 | 5.74±0.54 | 5.24±0.49 |
| 103.6-119.2 | 4.03±0.34 | 3.41±0.32 | 111.6-128.4 | 6.36±0.60 | 7.25±0.64 |
| 120.0-134.2 | 3.61±0.26 | 3.91±0.36 | 129.2-144.5 | 6.27±0.53 | 6.13±0.50 |
| 134.8-147.9 | 4.29±0.37 | 3.19±0.30 | 145.3-159.4 | 6.92±0.61 | 7.52±0.69 |
| 148.6-161.1 | 4.17±0.36 | 3.84±0.34 | 160.2-173.7 | 7.83±0.78 | 7.94±0.78 |

| A27I[%] ACE O E[MeV/n] | $A < 0$ 2007-2009 | $A > 0$ 2017-2019 |
|---|---|---|
| 59.0-75.6 | 3.14±0.19 | 3.67±0.29 |
| 77.2-103.8 | 2.76±0.26 | 2.94±0.24 |
| 105.1-127.2 | 3.19±0.25 | 2.85±0.24 |
| 128.3-147.8 | 4.04±0.40 | 3.59±0.29 |
| 148.7-166.5 | 3.03±0.21 | 3.49±0.28 |
| 167.4-183.8 | 3.55±0.33 | 4.08±0.35 |
| 184.7-200.4 | 4.22±0.32 | 3.96±0.35 |

**Table 4.** Amplitude of the 27-day variation of the GCR intensity for Neon, Silicon and Iron by *ACE/CRIS* for $A > 0$ and $A < 0$ polarities.

| A27I[%] ACE Ne E[MeV/n] | $A < 0$ 2007-2009 | $A > 0$ 2017-2019 | A27I[%] ACE Si E[MeV/n] | $A < 0$ 2007-2009 | $A > 0$ 2017-2019 |
|---|---|---|---|---|---|
| 69.4-89.0 | 7.66±0.55 | 6.87±0.58 | 86.3-110.9 | 7.97±0.64 | 6.44±0.60 |
| 91.0-122.5 | 6.23±0.60 | 5.89±0.47 | 113.3-153.1 | 5.59±0.53 | 5.11±0.36 |
| 124.0-150.3 | 6.84±0.54 | 6.72±0.67 | 155.0-188.3 | 4.93±0.46 | 5.78±0.55 |
| 151.6-174.9 | 7.32±0.66 | 6.24±0.48 | 190.0-219.6 | 6.39±0.52 | 5.96±0.52 |
| 176.0-197.3 | 7.35±0.61 | 7.51±0.61 | 221.0-248.3 | 6.87±0.57 | 6.04±0.60 |
| 198.3-218.0 | 8.69±0.89 | 8.07±0.63 | 249.5-274.9 | 8.24±0.59 | 8.95±0.68 |
| 219.1-237.9 | 8.40±0.84 | 8.27±0.66 | 276.3-300.5 | 7.89±0.72 | 7.97±0.63 |

| A27I[%] ACE Fe E[MeV/n] | $A < 0$ 2007-2009 | $A > 0$ 2017-2019 |
|---|---|---|
| 123.5-159.6 | 5.79±0.45 | 6.63±0.63 |
| 163.3-222.3 | 5.15±0.44 | 5.10±0.47 |
| 225.1-275.3 | 5.76±0.56 | 5.17±0.44 |
| 277.8-322.8 | 6.09±0.42 | 6.76±0.56 |
| 324.9-366.7 | 7.73±0.77 | 6.63±0.57 |
| 368.6-407.7 | 8.06±0.74 | 9.06±0.71 |
| 409.9-447.7 | 9.26±0.76 | 10.18±0.95 |

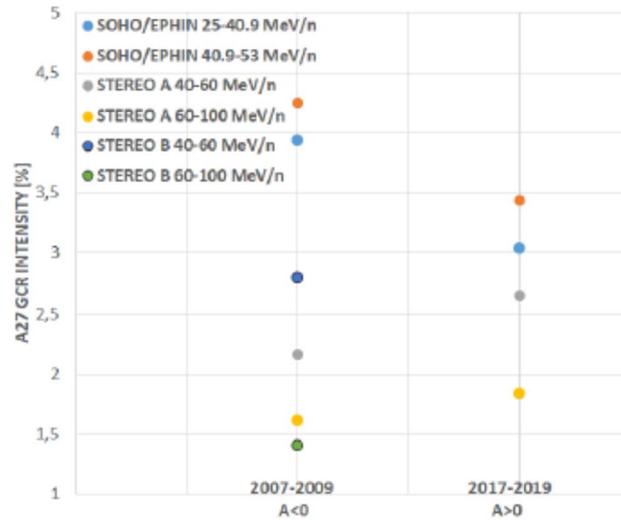

**Fig. 4.** The same as in Fig. 3 for the 27-day variation of the GCR proton intensity by *SOHO/EPHIN* and *STEREO* A, B.

The relationships in time-frequency domain around solar rotation period between GCR and other studied time series were analyzed using the cross wavelet transform (XWT). The first panel of Figure 7 shows cross-wavelet analysis results for the cosmic rays and sunspot number. The cosmic rays changes were mostly led by the SSN of _ 135-150 deg. This behaviour was much weaker in the SRF performance. The third panel of Figure 7 shows that SWs and GCR were in anti-phase in 06.2019-09.2019. During other time intervals where GCR and SWs had high common power: 02.2017-06.2017 or 09.2018-01.2019 SWs was leading GCR of _ 120-135 deg. Similar changes were observed in the case of SWT, but the opposite direction in phase distribution were in the case of SWd (Figure 7, the fourth panel). Tracing the time evolution of the high mutual power of HMF strength B and GCR (Figure 7, the second panel) they were in anti-phase in 05.2019-06.2019. In 02.2017-03.2017 and 10.2018-12.2018 GCR was led by B of _ 70-100 deg. There was an interval 01.2018-03.2018, during which B and GCR were in phase. Much stronger common power was observed between HMF components Bx; By and Bz and GCR.

The first panel of Figure 8 presents results of the coherency analyses between cosmic rays flux measured by Oulu NM and solar activity level expressed by its proxy, i.e. SSN. There are time inter-

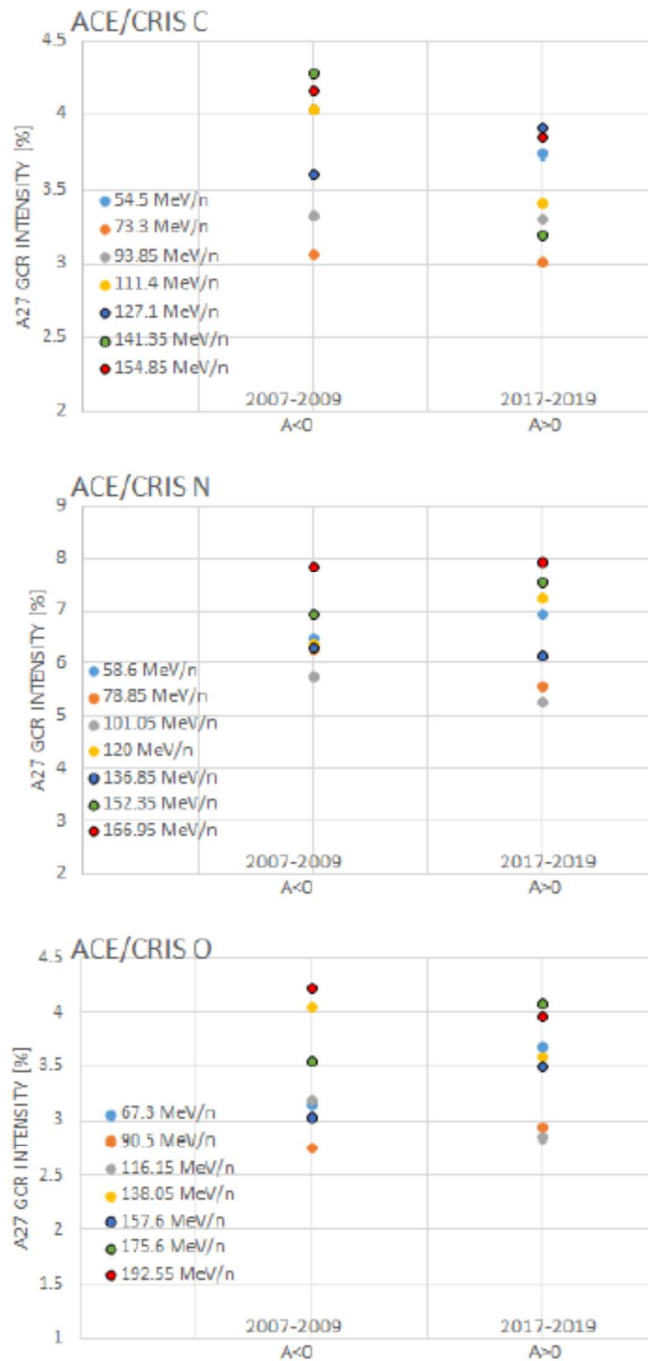

Fig. 5. The same as in Fig. 4 for GCR Carbon, Nitrogen and Oxygen by *ACE/CRIS*.

vals where SSN leads GCR averagely of _ 135 deg: 03.2017-05.2017, 06.2018-08.2018 10.2018-11.2018, 04.2019-06.2019 and 08.2019-11.2019. Similar, but weaker leading is also visible in SRF and GCR WTC. Figure 8, the third panel displays a significant coherence at the period of _ 27 days between cosmic rays and solar wind speed being almost in anti-phase from 02.2017 to 07.2017, from 09.2018 to 01.2019 and from 07.2019 to 09.2019. Practically the same situation is visible in the case of SWT, although between 02.2017 and 07.2017 SWT rather leads GCR of 90 deg. In the case of SWd (Figure 8, the fourth panel) and GCR they are rather in phase. Mutual behavior in time-frequency domain of cosmic rays and HMF strength B shows Figure 8, the second panel.

There is a period of a phase agreement from 10.2017 till 04.2018 visible also in Bx and By HMF components, but not in Bz. Later, during 09.2018 -01.2019 B is leading GCR of _ 150 deg and in 05.2019-09.2019 B and GCR are practically in anti-phase.

Usage of WTC can show an existence of a certain periodicity even if in one of considered time series this periodicity was rather weak, but in the second one there was observed a very strong signal. This is one of the reasons of confronting these results (Figure 8) with a wavelet coherency being a crosswavelet spectrum normalised to individual wavelet transform power spectrum (Figure 7). This method allows to observe a common power more clearly.

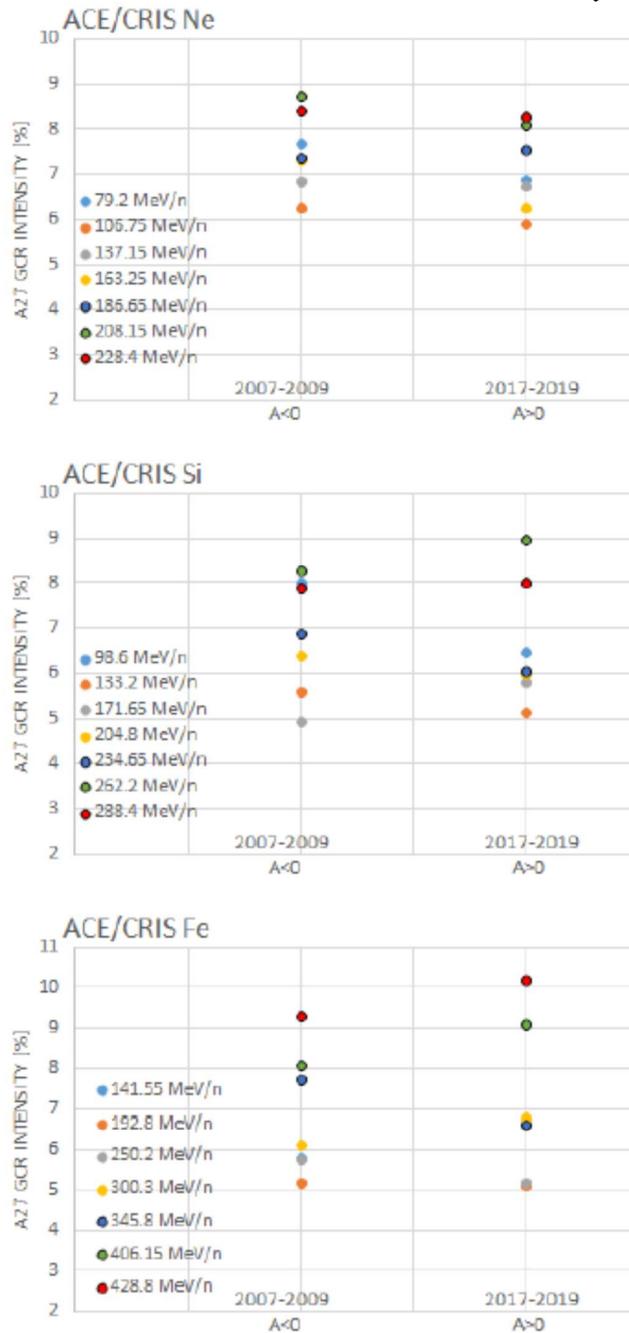

Fig. 6. The same as in Fig. 4 for GCR Neon, Silicon and Iron by *ACE/CRIS*.

**5. Discussion**

Up to now for explaining the polarity dependence of the 27-day variation of GCR several approaches were proposed, e.g., the polarity dependent diffusion coefficients (Richardson et al. 1999; Richardson 2004), heliolongitudinal asymmetry of the solar wind velocity (Modzelewska & Alania 2012) and combination of solar wind convection and drift effects (Gil & Mursula 2017). Richardson et al. (1999) suggested that the response of the cosmic rays to solar wind speed enhancements seems to be diminished in $A < 0$. Furthermore they interpreted it in the sense of A-dependence of transport coefficients (Chen & Bieber 1993), especially by diminishing of the radial diffusion coefficient which may increase the effect of solar wind convection during $A > 0$. It

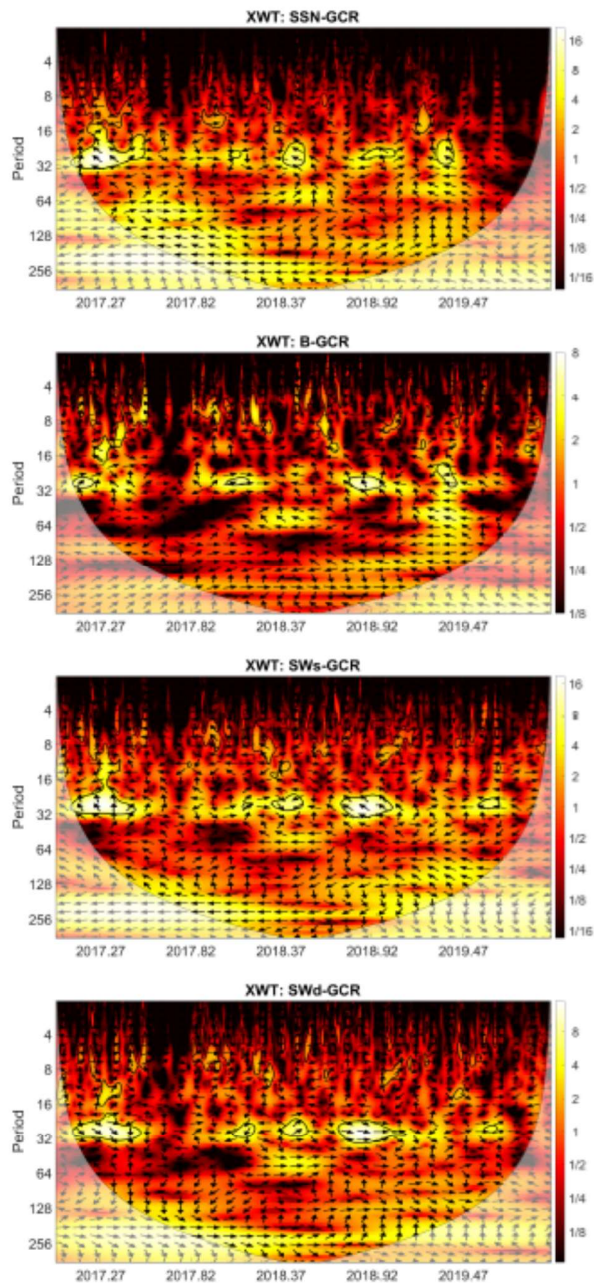

**Fig. 7.** The cross wavelet transform (XWT) between cosmic rays flux measured by Oulu NM and solar wind and solar parameters: (a) SSN, (b) SWs, (c) SWd and (d) $B$ in 2017-2019, where arrows represent the phase difference, the shaded zone is a cone of influence and the black ovals label 5% significance level

has to be underlined that Richardson et al. (1999) did not show any remarkable differences in solar wind structures between A > 0 and A < 0. In contrary (Modzelewska & Alania 2012; Alania et al. 2008) showed that the heliolongitudinal distribution of the phase of the 27-day variation of the solar wind velocity demonstrated that a long–lived (∼ 22 years) active heliolongitudes existing on the Sun preferentially for the A > 0, being the source of the long-lived 27-day variation of the solar wind velocity, and afterwards, it could be considered as the general source of the 27-day variations of the GCR intensity and anisotropy. Moreover, Modzelewska & Alania (2012) showed that the amplitudes of the 27-day variation of the solar wind velocity were about two times greater for the A > 0 epochs than for the A < 0. As a consequence, it was shown by (Alania et al. 2008; Gil et al. 2005; Modzelewska & Alania 2012), that recurrent changes of the solar wind velocity were crucial in the modulation mechanism of the 27-day variation of GCR in the minimum epoch of solar activity. Therefore in the sequence of papers (Alania et al. 2010, 2011; Modzelewska & Alania

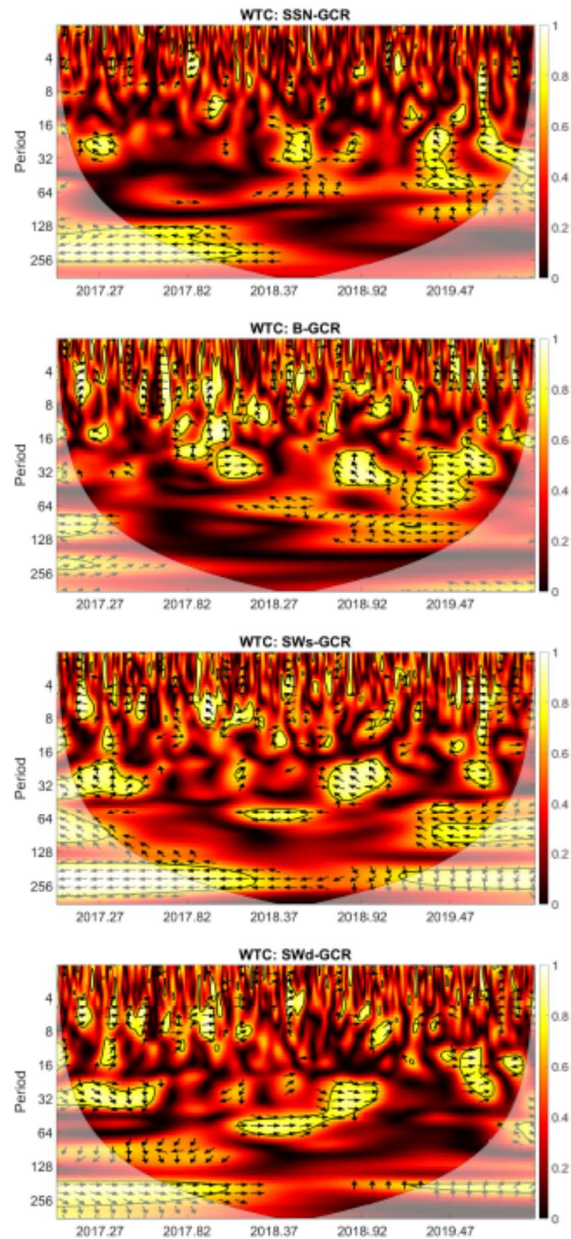

**Fig. 8.** The coherency analyses between cosmic rays flux measured by Oulu NM and solar wind and solar parameters: (a) SSN, (b) SWs, (c) SWd and (d) B in 2017-2019, where arrows represent the phase difference, the shaded zone is a cone of influence and the black ovals label 5% significance level

2013) studied theoretically successfully reproducing the 27-day variation of GCR observed by NMs during 2007-2008. This model was based on the Parker transport equation that incorporates the recurrent changes of solar wind speed and consistent divergence-free HMF with a full 3-d anisotropic diffusion tensor (Alania 1978, 2002). In contrary Kota and Jokipii (2001) in order to reproduce the polarity dependence of the 27-day variation used a non-stationary transport model that incorporated the southward displacement of the heliospheric current sheet (HCS) and CIRs effect.

Lately, the polarity dependence of the 27-day amplitude of GCR was confirmed experimentally by Gil & Mursula (2017), explaining it by combination of drift effect and solar wind convection. Gil & Mursula (2017) based on the computed amplitudes of the first two harmonics of the recurrent variation of GCR intensity during 1964-2016, showed for the first time that the mean values of amplitudes in the solar minima experience a clear declining trend. Gil & Mursula (2017) stated that it may be due to the weakening of polar magnetic fields during the previous four solar activity cycles (Smith & Balogh 2008) and the subsequent growth, in the heliolatitudes, of the range of the HCS (e.g. Virtanen & Mursula 2010). Hence, Earth have spent more time within the HCS in conditions of slow solar wind in the last solar minimum leading to a significant decrease in amplitudes of the both harmonics. Moreover, the relative damping of the second harmonic was stronger due to the fact that more probable was that Earth was outside the HCS once than twice or even more times during single solar rotation (Gil & Mursula 2017).

Recently, Leske et al. (2019) and Ghanbari et al. (2019) based on ACE data, compared the 27-day variations of GCRs in the solar minima 23/24 and 24/25. Leske et al. (2019) stated that the amplitude of the GCR variations is similar during the last two minima 23/24 (2007-2009) and 24/25 (2017-2019), nonetheless that of the anomalous cosmic ray (ACR) variations is substantially less in 2017-2019 than in the previous minimum 2007-2009. They proposed that the cause of the recurrent variations is not only the polarity of the HMF by the bidirectional latitudinal gradient near the HCS (even when the HCS is crossed more than ones per solar rotation, the clear 27-day variation still persists). Moreover, they found that the direction of HCS intersections is not a crucial factor, but it seems more important whether the observer changes position during HCS intersections, i.e. leaves the area under the influence of the dominant coronal hole or enters to it. Ghanbari et al. (2019) proposed that the convection of solar wind does not play a significant role in the vicinity of CIRs and indicated that the GCR intensity is inversely proportional to the perpendicular diffusion coefficient around CIR.

Engelbrecht & Wolmarans (2020) studying values for the several heliospheric turbulence quantities significant to GCR modulation such as, magnetic variances, squared HMF magnitudes, as well as parallel and perpendicular proton mean free path at Earth as a result of the cosmic ray long-term modulation model concluded that the diffusion and drift coefficients of GCR protons do not deviate much from each other during the previous minima. The need of inclusion of turbulence analysis when GCR characteristics are investigated was also underlined (Engelbrecht & Wolmarans 2020). Zhao et al. (2018) and Caballero-Lopez et al. (2019) using the spacecraft observations analyzed the turbulence properties vs solar cycle, and subsequently found the solar cycle dependence of cosmic-ray diffusion coefficients derived from the quasi linear and nonlinear theories.

This paper shows that the 27-day variations of the GCR anisotropy and intensity observed by NMs are larger in $A > 0$ than in $A < 0$ polarity. The same was recently reanalyzed for IMP8 for the previous solar cycles (Shen et al. 2020), conforming the results of (Richardson et al. 1999). This effect may be naturally linked to A-dependent drift effect in the global HMF. However, the 27-day variations of lower energies ($< 1$ GeV) GCR protons registered by STEREO A, B and SOHO/EPHIN and heavier species by ACE/CRIS observed in solar minima 23/24 and 24/25 remain at the same level and seem not to be polarity dependent. Solar minima 23/24 and 24/25 are characterized by the highest GCR intensity than in any previous $A > 0$ and $A < 0$. At the same time one can observe the diminishing trend in the 27-day amplitudes by IMP8 (Shen et al. 2020) and NMs (Gil & Mursula 2017), most probably due to weakening of solar polar fields. As a consequence, possibly, the GCR particle drifts may be suppressed by diffusion (turbulence) in background weakening B, this mechanism was earlier suggested by (e.g. Minnie et al. 2007). De Simone et al. (2011) found the value of bidirectional latitudinal gradient measured for low energy protons by Ulysses and PAMELA to be smaller than expected from theories.

Recurrent variations connected with the solar rotation for low energy ($< 1$ GeV) cosmic rays are more sensitive to the enhanced diffusion effects leading to the same level of the 27-day amplitudes for $A > 0$ and $A < 0$ polarities. High energy ($> 1$ GeV) cosmic rays observed by NMs, effective energy $10 \div 15$ GeV, are more sensitive to the large-scale drift effect resulting in the larger amplitudes of the 27-day GCR variations in the $A > 0$ polarity than in the $A < 0$. This is in an agreement with theoretical expectation of the drift reduction factor in terms of the rigidity due to turbulence by Engelbrecht et al. (2017). Moloto & Engelbrecht (2020) employing this reduction factor found that long-term GCR modulation effects can be explained by the solar-cycle dependent interplay between drift and diffusion, which were moderated by the solar cycle dependent influence of turbulence on the GCR drift coefficient.

In spite of the progress has been made from the experimental and theoretical point of view in understanding the modulation processes governing the 27-day variation of the GCR intensity and anisotropy, the problem is up to now not totally solved. Due to the complexity of GCR modulation around CIR more sophisticated numerical models (e.g. Luo et al. 2020; Guo & Florinsky 2016; Wiengarten et al. 2016; Wawrzynczak et al. 2015) should be tested on this problem in the future.

## 6. Conclusions

1. We confirm the polarity rule in the behaviour of the amplitudes of the 27-day variations of the GCR anisotropy and intensity observed by NMs in the solar minima: 23/24 (2007-2009) and 24/25 (2017-2019), namely larger amplitudes are observed for $A > 0$ polarity epoch.

2. The amplitudes of the 27-day variations of the GCR intensity observed by ACE/CRIS, STEREO A, B and SOHO/EPHIN, in the solar minima 23/24 and 24/25 remain at the same level and do not seem to be polarity dependent.

3. Recurrent variations connected with the solar rotation for low energy ($< 1$GeV) cosmic rays are more sensitive to the enhanced diffusion effects, leading to the same level of the 27-day amplitudes for $A > 0$ and $A < 0$ polarities. Whereas high energy ($> 1$GeV) cosmic rays observed by NMs, are more sensitive to the large-scale drift effect resulting in the 22-year Hale cycle of the 27-day GCR variations, with the larger amplitudes in the $A > 0$ polarity than in the $A < 0$. Nevertheless, processes around CIR are more complex and need further study, e.g. because of competition between modulation and acceleration of cosmic rays around stream interaction regions.

Acknowledgments:
Oulu neutron monitor count rates: http://cosmicrays.oulu.fi, SOHO/EPHIN, ACE/CRIS, heliospheric and solar wind parameters are from OMNI, STEREO A, B HET: http://www.srl.caltech.edu/STEREO/.
XWT & WTC tools by Grinsted et al.:
http://www.pol.ac.uk/home/research/waveletcoherence/.
We acknowledge the financial support by the Polish National Science Centre, grant number 2016/22/E/HS5/00406.

## References

Alania, M.V. 1978, Proc. of Institute of Geophys. Georgian Academy of Sci., 5, Tbilisi, Georgia (in Russian)
Alania, M.V., Baranov, D. G., Tyasto, M. I. et al. 2001, ASR, 27, 619
Alania, M.V. 2002 Acta Physica Polonica B, 33, 4, 1149
Alania, M.V., Gil, A., Iskra, K., et al. 2005, Proc. 29th ICRC,2, 215
Alania, M.V., Gil, A., & Modzelewska, R. 2008, ASR, 41, 280
Alania, M.V., Modzelewska, R., & Wawrzynczak, A. 2010, ASR, 45, 421
Alania, M.V., Modzelewska, R., & Wawrzynczak, A. 2011, Solar Phys., 270, 629
Bazilevskaya, G.A. 2000, SSR, 94, 25
Bazilevskaya, G.A., Broomhall, A.M., Elsworth, Y., et al. 2014, SSR, 186, 359
Caballero-Lopez, R.A.,. Engelbrecht, N.E., & Richardson, J.D. 2019, ApJ, 883, 73
Chowdhury, P., Kudela, K., & Moon Y. J. 2016, Solar Phys., 291, 581
Chen, J., & Bieber, J.W. 1993, ApJ, 405, 375
Decker, R.B., Krimigis, S.M., Ananth, A.G., et al. 1999, Proc. 26th ICRC,7, 512
Engelbrecht, N.E., Strauss, R.D., le Roux, J.A., et al. 2017, ApJ, 841, 107
Engelbrecht, N.E., & Wolmarans, C.P. 2020, ASR, 66, 2722
Ghanbari, K., Florinski, V., Guo, X., et al. 2019, ApJ, 882, 54
Gil, A. & Alania, M. V. 2001, Proc. 27th ICRC, 9, 3725
Gil, A., Iskra, K., Modzelewska, R., et al. 2005, ASR, 35, 687
Gil, A., & Mursula, K. 2017, A&A, 599, A112
Gil, A., & Mursula, K. 2018, JGR, 123, 6148
Grinsted, A., Moore, J.C., & Jevrejeva, S. 2004, Nonlin. Proc. Geophys., 11, 561
Guo, X., & Florinsky V. 2016, ApJ, 826, 13pp
Iskra, K., Alania, M.V., Gil, A., et al. 2004, Acta Physica Polonica B, 35, 1565
Jian, L.K., Luhmann, J.G., Russell, C.T. et al. 2019, Sol. Phys. 294, pp.31
Jevrejeva, S., Moore, J.C., & Grinsted, A. 2003, JGR: Atmospheres 108, 4677
Kota, J., & Jokipii, J.R. 2001, Proc. 27th ICRC,9, 3577
Kudela, K., & Sabbah, I. 2016, Sci. China Technological Sci., 59, 547
Leske, R., Cummings, A.C., Mewaldt, R.A., et al. 2011, Proc. 32nd ICRC,11, 194
Leske, R., Cummings, A.C., Mewaldt, R.A., et al. 2019, Proc. 36th ICRC, 1105
Lomb, N.R. 1976, Astrophys. Space Sci., 39, 447
Luo, X., Zhang, M., Feng, X., et al. 2020, ApJ, 899, 90
Mckibben, R.B., Simpson, J.A., Zhang, M., et al. 1995, SSR, 72, 403
Minnie, J., Bieber, J.W., Matthaeus, W.H., et al. 2007, ApJ, 670, 1149
Modzelewska, R., & Alania, M.V. 2012, ASR, 50, 716
Modzelewska, R., & Alania, M.V. 2013, Sol. Phys., 286, 593
Modzelewska, R., & Alania, M.V. 2018, A&A, 609, A32

Modzelewska, R., Alania, M.V., Gil, A., et al. 2006, Acta Phys. Pol. B, 37, 1641
Moloto, K.D., & Engelbrecht, N.E. 2020, ApJ, 894, 121
Potgieter, M.S. 2013, Living Reviews in Solar Physics, 10, 3
Richardson, I.G., Cane, H.V., & Wibberenz, G.A. 1999, JGR, 104, 12549
Richardson, I.G. 2004, SSR, 111, 267
Richardson, I.G. 2018, Living Rev. Solar Phys., 15, 1
Shen, Z., Qin, G., Zuo, P., et al. 2020, ApJ, 900, 2
de Simone, N., Di Felice, V., Gieseler, J., et al. 2011, Astrophys. Space Sci. Trans. 7, 425
Smith, E.J., & Balogh, A. 2008, GRL, 35, L22103
Torrence, C., & Webster, P.J. 1999, Journal of Climate, 12, 2679
Usoskin, I.G. 2017, Living Rev. Solar Phys., 14, 3
Vernova, E.S., Tyasto, M.I., Baranov, D.G., et al. 2003, ASR, 32, 621
Virtanen, I. I., & Mursula, K. 2010, JGR, 115, A09110
Wawrzynczak, A., Modzelewska, R. & Gil, A. 2015, J. Phys.: Conf. Ser., 574, 012078
Wiengarten, T., S. Oughton, S., Engelbrecht, N.E., et al. 2016, ApJ, 833, 17
Zhao, L.L., Adhikari, L., Zank, G.P., et al. 2018, ApJ, 856, 94